\documentclass[twocolumn,amssymb,amssymb,aps,pra]{revtex4} 
\usepackage{epsfig}
\usepackage{amsfonts}

\begin{document}

\title{Space-Time geometry and thermodynamic properties of a
self-gravitating ball of fluid in phase transition}

\author{Jos\'e D. Polanco}\email[e-mail: ]{pepediaz@ifi.unicamp.br}
\affiliation{Instituto de F\'{\i}sica `Gleb Wataghin', Universidade
Estadual de Campinas, 13083-970, Campinas, SP, Brasil}

\author{Patricio S. Letelier}\email[e-mail: ]{letelier@ime.unicamp.br}

\author{Maximiliano Ujevic}\email[e-mail: ]{mujevic@ime.unicamp.br}
\affiliation{Departamento de Matem\'atica Aplicada, Instituto de
Matem\'atica, Estat\'{\i}stica e Computa\c{c}\~ao Cient\'{\i}fica,
Universidade Estadual de Campinas, 13081-970, Campinas, SP, Brasil}

\begin{abstract}

A numerical solution of Einstein field equations for a spherical symmetric
and stationary system of identical and auto-gravitating particles in phase
transition is presented. The fluid possess a perfect fluid energy momentum
tensor, and the internal interactions of the system are represented by a
van der Walls like equation of state able to describe a first order phase
transition of the type gas-liquid. We find that the space-time curvature,
the radial component of the metric, and the pressure and density show
discontinuities in their radial derivatives in the phase coexistence
region. This region is found to be a spherical surface concentric with the
star and the system can be thought as a foliation of acronal, concentric
and isobaric surfaces in which the coexistence of phases occurs in only
one of these surfaces.  This kind of system can be used to represent a
star with a high energy density core and low energy density mantle in
hydrodynamic equilibrium.

\end{abstract}

\maketitle

\section{Introduction}

In relativistic astrophysics and pure general relativity the concept of
``phase transition" is usually associated with very different scenarios.
In general relativity we have the spherically symmetric collapse of a
massless scalar field or Yang-Mills field that leads to a formation of a
black hole \cite{cho,cho:chm}. In the astrophysical context we have first
order phase transition in neutron star models, where we found matter
condensation from neutron matter to pion-condensed matter and quark matter
\cite{kam,kam2,zdu:hae}.  Also we can have a phase transitions in the core
of a rotating neutron star \cite{hei:hjo}. In the last few years, the
phase transition phenomenon have become important in relativistic
astrophysics since we can have emission of gravitational waves
\cite{mar:vas,che:dai} associated to this transition.  Observational
evidence of this phenomenon can be attained \cite{gle} with the present
Laser Interferometer Gravitational-wave Observatory (LIGO). The LIGO-I
maximum distance for detection of gravitational waves is about 6.4 Mpc,
well beyond the Andromeda galaxy (M31), whereas the next generation
LIGO-II detectors will probably see phase transition events at distances
two times longer \cite{mar:vas}.

The model studied in this work is indirectly related to the topics mention
above. Our aim is to study the behavior of the thermodynamic properties
and the space-time structure in a simple albeit important astrophysical
object, a self-gravitating ball of fluid performing a generic first order
phase transition. In particular, the Riemann-Christoffel curvature tensor
is study in some detail for the space-time associated the different
regions of the matter that presents a first order phase transition.

To model this system in the context of General Relativity, we used the TOV
equations (Tolman-Oppenheimer-Volkoff)  \cite{tol,Oppi}. The TOV equations
represent the Einstein field equations for a stationary, spherical
symmetric system with a perfect fluid energy momentum tensor and a local
equation of state.

The most simple and general equation of state to model a phase transition
in a classical system of particles, is the van der Walls equation of state
together with the Maxwell energy balance construction. The Maxwell energy
balance construction have been studied in a rigorous treatment in systems
with two-body interaction potential proving the equal-area rule at the
transition region in the equation of state \cite{leb:pen}. Also the non
analyticity of the van der Walls equation of state together with the
Maxwell energy balance construction appears to be generic for first order
phase transitions originated in microscopic short range forces
\cite{fri:pfi,fri:pfi2}.

To solve the TOV equations one needs a smooth equation of state. In order
to overcame the fact that in a first order phase transition we will have a
non analytic equation of state we shall perform several smoothing
interpolations. Afterward we show that the real situation can be obtained
as a limit that is independent of the used interpolations.

The article is organized as follows, in section \ref{two} we present the
basic equations that model our system, in particular we study the TOV
equations and present a van der Walls like equation of state that models a
phase transition. In section \ref{three}, we analyze the special case of
constant pressure. This case needs to be treated in a different form
because, strictly, the TOV equations are not valid in this case. In
section \ref{four}, we present the behavior of the thermodynamic
variables, the metric components, the Kretschmann curvature invariant and
the components of the Riemann-Christoffel curvature tensor inside the star
undergoing a phase transition. Finally, in section \ref{five} we
summarized our results.

\section{Stellar model for a system in phase transition} \label{two}

We consider a static spherical symmetric space-time described by the line
element

\begin{equation}
ds^{2}= e^\nu c^{2}dt^{2}- e^\lambda dr^{2} - r^{2}(d\theta
^{2}+\sin ^{2}\theta d\varphi^{2}), \label{eq:ds2}
\end{equation}

\noindent where $\nu=\nu(r)$ and $\lambda= \lambda(r)$. The matter is
represented by the perfect fluid energy momentum tensor

\begin{equation}
T_{\alpha \beta }=\rho \,u_{\alpha }u_{\beta }+ \frac{p}{c^{2}}
(u_{\alpha }u_{\beta }-g_ {\alpha \beta } ), \label{eq:Talfabeta}
\end{equation}

\noindent where $\rho$ is the energy density, $u^\alpha$ is the
four-velocity, $c$ is the speed of light, and $p$ the pressure measured in
the local rest frame. We also assume that the pressure and the energy
density are related by an equation of state of the form $p=p(\rho)$.

The Einstein field equations,

\begin{equation}
R_{\alpha \beta }=-\zeta\left( T_{\alpha \beta }-\frac{1}{2}
g_{\alpha \beta }T\right),  \label{eq:EEInstein}
\end{equation}

\noindent where $\zeta=\frac{8\pi G}{c^{2}}$ and $G$ is the gravitational
constant, can be written with the help of (\ref{eq:ds2}) and
(\ref{eq:Talfabeta}) as a set of three equations,

\begin{eqnarray}
&& e^{-\lambda} \left[ -\frac{1}{2}\nu^{\prime\prime} + \frac{1}{4}\lambda 
^{\prime}\nu^{\prime}-\frac{1}{4} (\nu^{\prime})^2 - \frac{1}{r}\nu 
^{{\prime}}\right] =-\zeta 
\left( \frac{\rho }{2}+\frac{3}{2} \frac{p}{c^{2}}\right), \nonumber \\
&& e^{-\lambda} \left[ \frac{1}{2}\nu ^{{\prime}{\prime}
}-\frac{1}{4}\lambda ^{{\prime}}\nu ^{{\prime}}+\frac{1}{4} ( \nu ^{
{\prime}} ) ^{2}-\frac{1}{r}\lambda ^{{\prime}}\right] = -\zeta 
\left( \frac{\rho }{2}-\frac{1}{2}\frac{p}{c^{2}}\right), \nonumber \\
&& e^{-\lambda} \left[ \frac{1}{r^{2}}+\frac{1}{2r} ( 
\nu^{{\prime}}-\lambda ^{{\prime}} ) \right] -\frac{1}{r^{2}}=-\zeta 
\left( \frac{\rho }{2}-\frac{1}{2}\frac{p}{c^{2}}\right), 
\label{eq:14.12a} 
\end{eqnarray}

\noindent in which $^\prime$ denotes differentiation with respect to the
coordinate $r$. To write the equations above in a simpler form we use the
function $m(r)$ defined by,

\begin{equation}
m(r)=\frac{r}{2}(1-e^{-\lambda}). \label{eq:1-2m/r}
\end{equation}

$m(r)=G M(r)/c^2$ is the geometric mass which has dimensions of distance
and $M(r)$ is the quantity of mass inside a sphere of radius $r$. With
this change of variable, we obtain the system of equations known as the
TOV equations,

\begin{eqnarray}
&&p=p(\rho), \label{eq:E1} \\
&&m^\prime(r)=\frac{4\pi G}{c^2} r^2 \rho, \label{eq:E2} \\
&&p^\prime=-\frac{(\rho +\frac{p}{c^2} ) (m+\frac{4\pi G}{c^4} r^3 
p) c^2} {r ( r-2m ) }, \label{eq:E3} \\
&&e^{-\lambda} =1-\frac{2m(r)}{r}, \label{eq:E4} \\
&&\nu^{\prime}= -\frac{2p^{\prime}}{\rho c^2+p}. \label{eq:E5}
\end{eqnarray}

\noindent The first three equations allow us to solve for the geometric
mass $m(r)$, the pressure $p(r)$, and the energy density $\rho(r)$, i.e.
the thermodynamic quantities of the system. The last two equations define
the space-time geometry through the functions $\lambda(r)$ and $\nu(r)$
present in the metric.

\begin{figure*}
\epsfig{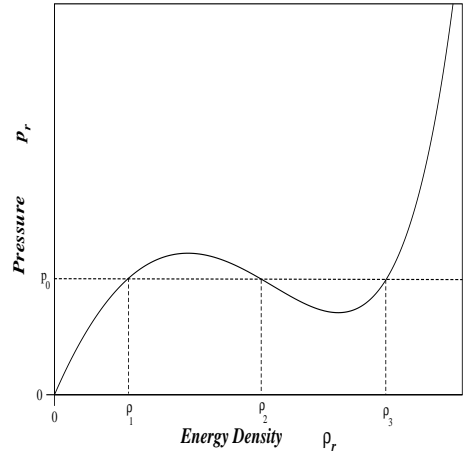} 
\caption{A typical van der Walls isothermal for a system in gas-liquid
phase transition using reduce quantities. The coexistence pressure $p_0$
and the coexistence region $\rho_1 \leq \rho_r \leq \rho_3$ is calculated
using the Maxwell construction.}
\label{fig:eevw}
\end{figure*}

The equation of state (\ref{eq:E1}) depends on the particle interactions
of the system. An equation of state that allows a phase transition is a
van der Walls like equation of state of the the form,

\begin{equation}
p=\frac{kT\rho}{1-b\rho}-a \rho ^2 ,  \label{eq:pvdW}
\end{equation}

\noindent where $a$ and $b$ are characteristic parameters of a particular
physical system. To solve the TOV equations using the equation of state
(\ref{eq:pvdW}), we have to specify the type of particle considered
through the parameters $a$ and $b$. Using the law of the corresponding
states is possible to obtain an equation of state which is independent of
these parameters, in other words, valid for any ``real gas". This equation
of state in the so called reduce quantities ($p_r$, $T_r$, $\rho_r$) is

\begin{equation}
p_{r}=\frac{8T_{r}\rho_{r}}{3-\rho_{r}}-3 {\rho_{r}}^2 ,
\label{eq:vander}
\end{equation}

\noindent with $p_{r}$, $T_{r}$ and $\rho_{r}$ defined by the adimensional
relations

\begin{equation}
p_{r}=\frac{p}{p_{c}},\hspace{1cm} \rho_{r}=\frac{\rho}{\rho_{c}}
,\hspace{1cm}T_{r}=\frac{T}{T_{c}} ,
\end{equation}

\noindent where $(p_c,\rho_c,T_c)= \left(\frac{a}{27b^2}, \frac{1}{3b},
\frac{8a}{27kb}\right)$ are the values of the pressure, energy density and
temperature in which the first and second derivative of (\ref{eq:vander})
become zero simultaneously. Also $0\leq \rho_r<3$. In Fig. \ref{fig:eevw}
we show a typical van der Walls like isothermal for a system in phase
transition (gas-liquid) in reduce coordinates. We know that the van der
Walls like equation of state describes well the regions of gas and liquid
but in the region of coexistence it presents physical inconsistencies,
e.g.  in the transition region we do not have dynamical equilibrium
between the phases. Therefore, we modify the equation of state
(\ref{eq:vander}) using the Maxwell energy balance construction
\cite{GREINER}. The Maxwell energy balance construction allows us to
calculate the pressure of coexistence $p_0$ and the coexistence energy
density interval [$\rho_1,\rho_3$] through the system of equations,

\begin{eqnarray}
&&\int\limits_{\rho_{1}}^{\rho_{3}}p_{r}\left(
\rho_{r} \right) d\rho_{r}= p_{0}\left( \rho_{3}-\rho_{1}\right),
\nonumber \\
&& \label{eq:qq1} \\
&&\frac{8T_{r}\rho_{1}}{3-\rho_{1}}-3 {\rho_{1}}^2 =p_{0}=
\frac{8T_{r}\rho_{3}}{3-\rho_{3}}-3 {\rho_{3}}^2 . \nonumber
\end{eqnarray}

\noindent First we fix a temperature value $T_r$ and then we solve the
system of equations (\ref{eq:qq1}) ( e.g.  with the Newton-Raphson method
for nonlinear systems). Once $p_0$ is found, we can say that the state
equations for our system in phase transition between gas and liquid is of
the form,

\begin{equation}
p_{r}=\left\{ 
\begin{array}{ccc}
\frac{8T_{r}\rho_{r}}{3-\rho_{r}}-3 {\rho_{r}}^2 & 0 \leq \rho{_r} 
 \leq \rho_1& 
({\rm gas})\\ 
p_{0}&\rho_1 \leq \rho{_r} \leq \rho_3 & ({\rm coexistence}) \\
\frac{8T_{r}\rho_{r}}{3-\rho_{r}}-3 {\rho_{r}}^2 & 
\rho_3 \leq \rho{_r}<3 & ({\rm liquid})
\end{array}
\right. . \label{eq:p(r)}
\end{equation}

We note that the equation of state (\ref{eq:p(r)}) is not of class $C^2$
because its derivatives have discontinuities in $\rho=\rho_1$ and
$\rho=\rho_3$. So, the TOV equations can not be solved around these energy
densities. Furthermore, in the coexistence region, the equation of state
satisfies the equation $p_{r}^{\prime}=0$ that brings no physical
solutions, e.g. negative masses, see equation (\ref{eq:E3}). For this
reason, the case of constant pressure has to be study independently. The
problem of smoothness above mentioned can be solved using an equation of
state of class $C^2$ or smoother instead of (\ref{eq:p(r)}).  This is done
redefining the equation of state (\ref{eq:p(r)}) in the region of
coexistence using a virial like equation to make, at least, the joints in
$\rho_1$ and $\rho_3$ of class $C^2$.  The line $p_r ={\rm constant}$ in
the coexistence region can be approximated by a straight line with a small
slope angle ($\alpha$). In this way, our equation of state for the system
becomes

\begin{equation}
p_r =\left\{ 
\begin{array}{cc}
\frac{8T_r \rho_r}{3-\rho_r}-3 {\rho_r}^2 & 0 \leq \rho_r 
\leq \rho_{1}-\chi \\
\sum\limits_{n=0}^{5} a_{n}\rho_r^n&\rho_{1}-\chi \leq 
\rho_r \leq \rho_{1}+\chi \\   
\sum\limits_{n=0}^{1} b_{n}\rho_r^n&\rho_{1}+\chi 
\leq \rho_r \leq \rho_{3}-\chi\\
\sum\limits_{n=0}^{5} c_{n}\rho_r^n&\rho_{3}-\chi \leq \rho_r 
\leq \rho_{3}+\chi\\
\frac{8T_r \rho_r}{3-\rho_r}-3 {\rho_r}^2 & 
\rho_{3}+\chi \leq \rho_r<3 
\end{array}
\right., \label{smootheqst}
\end{equation}

\noindent where $a_n$, $b_n$, $c_n$ are constants to be determined, and $2
\chi$ is the interval for the coupling polynomials. In our case we define
the coefficients $b_n$ so that the straight line pass through the point
$p_0$ at the middle of the interval [$\rho_1+\chi,\rho_3-\chi$], in this
way

\begin{equation}
b_0=p_{0}-\left(\frac{\rho_{3}+\rho_{1}}{\rho_{3}-\rho_{1}-2\chi}\right)\xi, 
\hspace{0.5cm} b_1=\frac{2\xi}{\rho_{3}-\rho_{1}-2\chi}, \label{bn}
\end{equation}

\noindent where $\xi \ll 1$ is directly associated with the angle $\alpha$
[$\tan \alpha = 2 \xi / (\rho_3-\rho_1 - 2 \chi$)] of the straight line in
the transition region. The polynomial coefficients $a_n$ and $c_n$ are
calculated using the six boundary conditions given by the continuity of
the pressure, its first derivative, and its second derivatives at the
junctions. We found that our $C^2$ class curve defined in this way is a
good approximation for the case of a gas-liquid transition. The analysis
of the case with $p={\rm constant}$ will be discussed in section
\ref{three}.

For generality sake, we want that the system of equations
(\ref{eq:E1}-\ref{eq:E5})  be independent on the characteristic parameters
of the gas. We introduce new variables $\widehat{r}$ and $\widehat{F}$
define as

\begin{equation}
r=\Lambda \widehat{r},\hspace{0.5cm} 
m=\Lambda \widehat{F}\widehat{r}^2,
\end{equation}

\noindent where the quantity $\Lambda$ is a scale factor.  Furthermore we
fix our units by choosing

\begin{equation}
\frac{a\rho _{c}}{c^{2}}=1 , \hspace{0.5cm} \frac{4\pi\rho_c G 
}{c^2} \Lambda^{2}\equiv 1 \label{eq:RR0} .
\end{equation}

\noindent Substituting these expressions in equations (\ref{eq:E1}),
(\ref{eq:E2}) and (\ref{eq:E3}), the internal structure of the star
undergoing a phase transition is found solving the system of equations

\begin{eqnarray}
&&\frac{\partial \rho_r }{\partial \widehat{r}}=-\frac{1}{(1-2\ 
\widehat{r}\widehat{F}) }
\left[ \widehat{F} +\frac{\widehat{r} p_r}{3}\right] 
\left[ 3\rho_r+p_r\right] 
\left[ \frac{\partial p_r}{\partial \rho_r }\right] ^{-1} \nonumber
\\
&&\frac{\partial \widehat{F} }{\partial \,\,\widehat{r}}=
\rho_r -2\frac{\widehat{F} }{\widehat{r}} , \label{eq:sis}
\end{eqnarray}

\noindent where for each region of the star we have to use the isothermal
corresponding to the temperature that allows a phase transition.

\section{Analysis of the case of constant pressure} \label{three}

 To perform the integration of the TOV equations through the
discontinuities of the coexistence region, we use the class $C^2$
equation of state (16) with the constants of the coupling polynomials
calculated using the six boundary conditions mentioned  in section \ref{two}.
The coexistence region is modeled, in first approximation, by a straight
line with a small slope angle ($\alpha \approx 10^{-3}$). Then  we
let the  angle $\alpha$  go to
zero to assure that the abrupt change presented in the energy density
figure (and other figures that depend on the energy density) are indeed
vertical lines. The same result can be obtained in  a different way.
Knowing the pressure and energy densities interval for the coexistence
region (e. g., from the Maxwell construction)  one can solved the TOV equation
independently for the liquid  phase as well as for and gas phase. Then the
 end points are joit 
with a straight line. With this second method we never integrate the TOV
equations through the discontinuities. Both methods
give the same results,  we use the first one because in our opinion the
results appear in a more transparent way.

To determine the behavior of the energy density in the coexistence region
and the radial interval in which the transition occurs in the limit
$\alpha \rightarrow 0$, we define a straight line passing through the
points $(\rho_i ,p_0)$ and $(\rho_f,p_0+\epsilon)$, where we see that for
a small $\epsilon$ [$\epsilon = (\rho_f-\rho_i) \tan(\alpha)$] a good
approximation of the line $p_r = p_0 = {\rm constant}$ is found. We choose
arbitrary values for the energy density interval $(\rho_i,\rho_f)=(1,2)$
and pressure $p_0=1$, with these values the equation of state in the
coexistence region is of the form,

\begin{equation}
\begin{array}{cc}
p_r=1+\epsilon (\rho_r -1)   \hspace{1cm}  &     1\leq \rho_r \leq 2
\end{array}.\label{eq:pconstante}
\end{equation}

\begin{figure*}
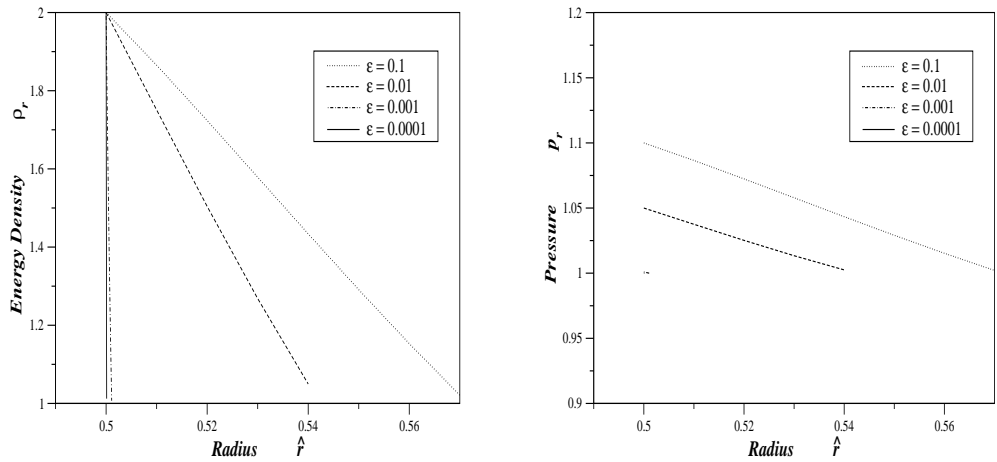

\epsfig{width=6cm,height=6cm,file=figure2a.eps}\hspace{1cm}
\epsfig{width=6cm,height=6cm,file=figure2b.eps} 
\caption{Analysis of the energy density and pressure profiles. We see from
the graphs that in the limit when $\epsilon \rightarrow 0$ ($p_r={\rm
constant})$, the energy density profile tends to a vertical line at a fixed
radius while the radial domain of the pressure profile becomes a point.}  
\label{fig:rocon}
\end{figure*}

\noindent When $\epsilon \rightarrow 0$ we obtain the case of constant
pressure $p_r=1$. In figure \ref{fig:rocon} we solved numerically the TOV
equations (\ref{eq:sis}) with the equation of state (\ref{eq:pconstante})  
for different values of $\epsilon$. In the limit $\epsilon \rightarrow 0$
we can affirm that the transition occurs only in a concentric spherical
surface of the star. This means that in the limit of $\epsilon \rightarrow
0$ for a given equation of state of the type (\ref{smootheqst}) with
(\ref{eq:pconstante}), the TOV system brings us a solution of a star model
that consists in a foliation of spherical, isobaric, and acronal surfaces.
This is in accordance with the stationary hypothesis made for the model,
because if the transition take place in a radial interval instead of a
fixed radius, the different gravitational forces at different radius will
make a non stationary system.

\section{Solution of TOV equations for a system in phase transition}  
\label{four}

\subsection{Thermodynamic properties of the star}

\begin{figure*}
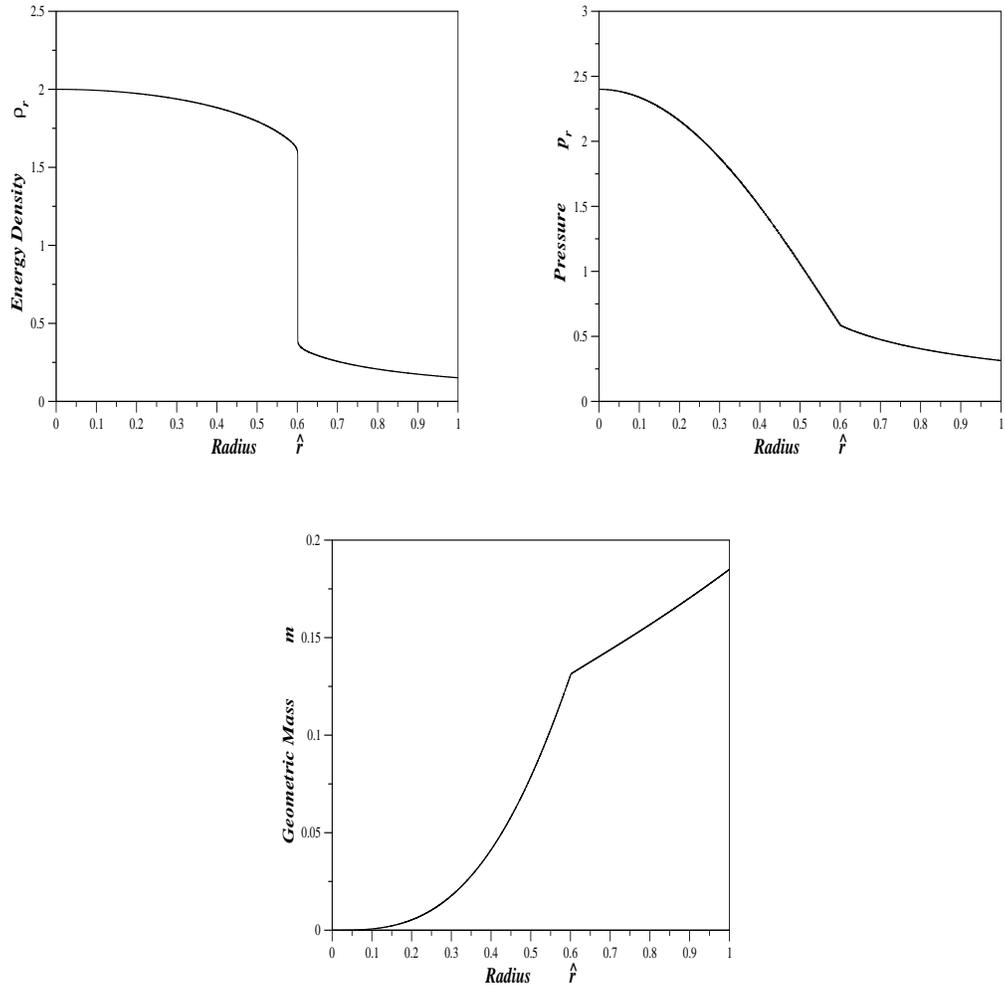
 
\epsfig{width=6cm, height=6cm, file=figure3a.eps} \hspace{1cm}
\epsfig{width=6cm, height=6cm, file=figure3b.eps}

\vspace{1cm} \epsfig{width=6cm, height=6cm, file=figure3c.eps}
\caption{Energy density, pressure and geometric mass profiles for the
interior of a star in phase transition. The discontinuities in their 
radial derivatives at $\widehat{r} \approx 0.6$ represent the change of 
phases from a liquid core (high energy density) to a gas mantle (low 
energy density).}
\label{fig:D(R)}
\end{figure*}

For didactic purposes we choose the isothermal with reduced temperature
$T_r=0.9$ and parameters ($\chi=0.005$, $\xi=0.00005$) to study the
thermodynamic properties and the space-time behavior. It can be verified
that all the isotherms that allow a phase transition have the same
qualitative behavior. With the values of the coexistence pressure $p_0 =
0.583$ and the energy density coexistence interval $(\rho_1,\rho_3) =
(0.350,1.623)$ found from (\ref{eq:qq1}), we can solve numerically the
system of equations (\ref{eq:sis}). The inner boundary condition can be
set recalling that $m(r)$ is proportional to the mass inside the sphere of
radius $r$, $m(r=0)=0$. So, the inner boundary condition for the new
variables are $\left(\widehat{r},\rho_{0}, \widehat{F}_{0}\right)=
\left(0,2,0\right)$; which represent a liquid core of high energy density,
the value $\rho_0=2$ is set arbitrarily. In figure \ref{fig:D(R)}, we show
the numerical solution of the TOV equations for the thermodynamic
quantities in the star with phase transition. The phase transition shows
up like a suddenly brake in the energy density at $r \approx 0.6 \Lambda$.  
Looking at the energy density plot we can think the high dense region with
$r \lesssim 0.6 \Lambda$ as a liquid core and the region with small
density $r \gtrsim 0.6 \Lambda$ as a gas mantle. The pressure has a
discontinuity in its radial derivative at $r \approx 0.6 \Lambda$ and
presents the expected behavior, i.e. high and low pressures appears in the
high energy density (core) an low energy density (mantle)  regions
respectively.  The geometric mass increases rapidly in the liquid region
and slowly in the gas region, it also presents a discontinuity in its
radial derivative at the same radius of the star.  We note that the energy
density and the pressure goes to zero for large values of $\widehat{r}$,
so in principle we can match continuously to a Schwarzschild free-space
solution, as was done in neutron-star models \cite{neutron}, or to an
exterior dust fluid (with zero pressure).

\subsection{Space-time geometry in a star in  phase transition}

The metric components in the star are found through the functions
($\nu$,$\lambda$) from equations (\ref{eq:E4}) and (\ref{eq:E5}). The
integration constant from equation (\ref{eq:E5}) can be obtained by
imposing that the component $g_{00}$ at a distant radius has to be equal
to the Schwarzschild exterior solution

\begin{equation}
g_{00}(r_s)=1-{\frac{2 m(r_s)}{r_s}},
\end{equation}

\noindent where $r_s$ is the radius of the star. The component $g_{11}$ is
calculated directly from equation (\ref{eq:E4}).

These components are depicted in figure \ref{fig:g00g11}. We note in this
figure that the component $g_{00}$ of the metric does not present
discontinuities in the phase transition region as found in the radial
derivative of the $g_{11}$ component at $r \approx 0.6 \Lambda$. This
motivate us to think that, in general, the phase transition could be
manifested only in the spatial part of the metric.

\subsection{Curvature associated with the phase transition}

To study the space-time curvature in a system with phase transition, we
used the Kretschmann invariant ($K=R^{\alpha \beta \gamma \delta}
R_{\alpha \beta \gamma \delta}$). With the help of relations
(\ref{eq:14.12a}), we find that the Kretschamnn invariant can be written
as

\begin{widetext}
\begin{equation}
K=\frac{4}{\Lambda^{4}}\left(\left(\rho_r+\frac{p_r}{3}-2
\frac{\widehat{F}}{r}\right)^2+
2\left(\rho_r-\frac{\widehat{F}}{r}\right)^2+
2\left(\frac{p_r}{3}+\frac{\widehat{F}}{r}\right)^2+
4\left(\frac{\widehat{F}}{r}\right)^2\right) .
\end{equation}
\end{widetext}

\begin{figure*}
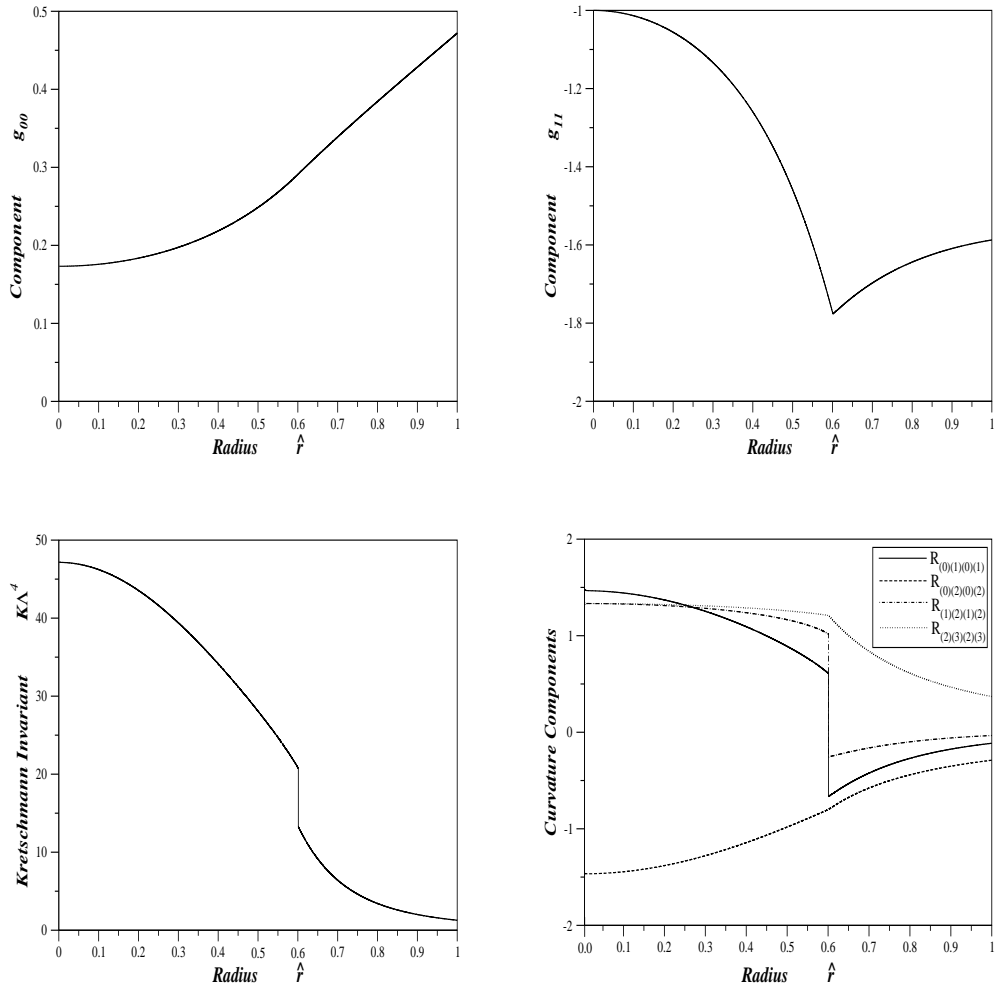

\epsfig{width=6cm, height=6cm, file=figure4a.eps}\hspace{1cm}
\epsfig{width=6cm, height=6cm, file=figure4b.eps}

\vspace{1cm} \epsfig{width=6cm, height=6cm, file=figure4c.eps}\hspace{1cm}
\epsfig{width=6cm, height=6cm, file=figure4d.eps} 
\caption{Behavior of the $g_{00}$ and $g_{11}$ components of the metric, 
the Kretschmann curvature invariant and the  components of the Riemann-Christoffel tensor in a star 
in  phase transition. We note that the phase transition is manifested 
only in the spatial component of the metric. The Kretschmann invariant, and
curvature components  $R_{(1)(2)(1)(2)}$ and  $R_{(0)(1)(0)(1)}$ 
experience a sudden change, this can be 
interpreted as a curvature transition inside the star due to the loss of 
energy needed for the phase transition to occur.}
\label{fig:g00g11} 
\end{figure*}

In figure \ref{fig:g00g11} we see that the Kretschmann invariant present a
similar behavior than the energy density, i.e.  highest curvature values
in the core of the star and at long distances the curvature tends to zero.  
The Kretschmann invariant depends on the values of the pressure, energy
density and geometric mass, so it is normal to expect a discontinuity in
the coexistence region at the same sphere radius ($r \approx 0.6
\Lambda$). For a better understanding of the curvature associated to the
phase transition, we project the Riemann tensor along the orthonormal
tetrad $\xi^\alpha_{(a)}$, where $\alpha$ is the contravariant tensor
index and $(a)$ a label distinguishing the particular vector. So, we have

\begin{equation}
R_{(a)(b)(c)(d)}=R_{\alpha \beta \gamma \delta} \xi^\alpha_{(a)} 
\xi^\beta_{(b)} \xi^\gamma_{(c)} \xi^\delta_{(d)},
\end{equation}

\noindent in our case the orthonormal tetrad is

\begin{eqnarray}
\xi^\alpha_{(0)}&=&(e^{-\nu/2},0,0,0), \nonumber \\
\xi^\alpha_{(1)}&=&(0,e^{-\lambda/2},0,0), \nonumber \\
\xi^\alpha_{(2)}&=&(0,0,r^{-1},0), \nonumber \\
\xi^\alpha_{(3)}&=&(0,0,0,[r \sin(\theta)]^{-1}),
\end{eqnarray}

\noindent and we obtain the following non null components of the Riemann
tensor:

\begin{eqnarray}
&& R_{(0)(1)(0)(1)} = \frac{1}{\Lambda^{2}}\left( \rho_r
+\frac{p_r}{3} -\frac{2\widehat{F}} {\widehat{r}}\right), \nonumber 
\\
&& R_{(0)(3)(0)(3)}=R_{(0)(2)(0)(2)} = -\frac{1}{\Lambda^{2}}\left(
\frac{\widehat{F}}{\widehat{r}}+\frac{p_r}{3}\right), \nonumber \\
&&R_{(1)(3)(1)(3)}=R_{(1)(2)(1)(2)} = \frac{1}{\Lambda^{2}}\left(\rho_{r}
-\frac{\widehat{F}}{\widehat{r}}\right), \nonumber \\ 
&& R_{(2)(3)(2)(3)} = 
\frac{1}{\Lambda^{2}}\left(\frac{2\widehat{F}}{\widehat{r}}\right). 
\label{riemanncompts}
\end{eqnarray}

\noindent These components are plotted in figure \ref{fig:g00g11}. We see
that all the components are functions of the class {\it C$^0$ }. We have
discontinuous radial derivative at $r \approx 0.6 \Lambda$.  In particular
the components $R_{(1)(2)(1)(2)}$ and $R_{(0)(1)(0)(1)}$ present the same
abrupt change that the Kretschmann scalar. The Kretschmann scalar can be
written as a function of the non zero components of the
Riemann-Christoffel curvature tensor (\ref{riemanncompts}). So, we can
assure that the abrupt change in the Kretschmann scalar is due to the
components $R_{(1)(2)(1)(2)}$ and $R_{(0)(1)(0)(1)}$. These two components
depend on the reduce energy density $\rho_r$. So, we can state that the
abrupt change in the Kretschmann scalar can be thought as a curvature
transition in the star due to the loss of energy needed for the phase
transition to occur, see the energy density profile from figure
\ref{fig:D(R)}.  We note that discontinuities in the scalar   curvature 
are not rare phenomena in nature. One simple example of this discontinuity
is provided by   a simple star model described in its exterior by the Schwarzschild solution and its interior by the solutions corresponding to a homogeneous sphere \cite{syn}. In this case there
exist a discontinuity in the  
 scalar  curvature mainly because the energy

density has a discontinuity, constant inside the star and zero outside.

\section{Conclusions} \label{five}

We studied a model of a stationary spherical symmetric system in phase
transition. The system is a ball of fluid made with identical and
auto-gravitating particles with the energy-momentum tensor of a perfect
fluid that obeys a van der Walls like equation of state. This model
appears to be the most simple and general model to describe a stellar
object that undergoes a phase transition. The phase transition in the
thermodynamic quantities is manifested as an abrupt change in the energy
density, and discontinuities in the radial derivatives of the pressure and
geometric mass at a fix radius. This radius divides the star in two
regions: one with a high energy density (a liquid core) and the other with
low energy density (a gas mantle). In the space-time geometry of the star,
we see an abrupt change in the Kretschmann invariant and some components
of the Riemann tensor. These can be thought as a curvature transition
process due to the loss of energy needed for the phase transition to
occur. A discontinuity in the radial derivative is also present in the
spatial component of the metric but not in its time component, this lead
us to think that phase transitions could be manifested only in the spatial
part of the metric. For the results obtained in this work, we interpret
the stationary system studied as a foliation of concentric and isobaric
spherical surfaces where the phase transition occurs in only one of these
surfaces. We note that the procedure used in this article can be applied
to other models, such as systems with different kind of particles and
systems characterized with more than one phase transition.

\section{Acknowledgment}

J.D.P. thanks CAPES for financial support; P.S.L. and M.U. thank FAPESP
for financial support, P.S.L. also thanks CNPq for financial support and
Dr. B. Boisseau for early discussions on the subjects of this work.

\end{document}